\documentclass[12pt,thmsa]{article}
\usepackage{sw20elba}
\usepackage{amsmath}
\usepackage{amsfonts}
\usepackage{amssymb}
\usepackage{graphicx}
\begin{document}

\author{ }

\begin{center}
{\Large Dependence of the superconducting transition temperature of\ MgB}%
$_{2}$ {\Large on pressure to 20 GPa\ }\vspace{0.5cm}

S. Deemyad,$^{a}$ J. S. Schilling,$^{a}$ J. D. Jorgensen,$^{b}$ and D. G.
Hinks$^{b}$\vspace{0.2cm}

$^{a}$\textit{Department of Physics, Washington University}

\textit{C.B. 1105, One Brookings Dr., St. Louis, MO 63130}

$^{b}$\textit{Materials Science Division, Argonne National Laboratory}

\textit{9700 South Cass Avenue, Argonne, IL 60439}\vspace{1cm}
\end{center}

\textbf{\noindent Abstract\vspace{0.2cm}}

The dependence of $T_{c}$ on nearly hydrostatic pressure has been measured for
an isotopically pure ($^{11}$B) MgB$_{2}$ sample in a helium-loaded
diamond-anvil-cell to nearly 20 GPa. \ $T_{c}$ decreases monotonically with
pressure from 39.1 K at ambient pressure to 20.9 K at 19.2 GPa. \ The initial
dependence is the same as\ that obtained earlier ($dT_{c}/dP\simeq-1.11(2)$
K/GPa) on the same sample in a He-gas apparatus to 0.7 GPa. \ The observed
pressure dependence $T_{c}(P)$ to 20 GPa can be readily described in terms of
simple lattice stiffening within standard phonon-mediated BCS
superconductivity.\vspace{1cm}

\section{INTRODUCTION}

The recent discovery \cite{n4} of superconductivity at high temperatures (40
K) in the $s,p$ metal compound MgB$_{2}$ was quite unexpected. \ To aid in the
search for even higher values of $T_{c}$ in this class of superconductor, one
would like to pinpoint both the pairing mechanism and the electronic/phononic
features which lend this simple binary compound its extraordinarily high value
of $T_{c}.$ \ Of particular value in this regard are studies of the variation
of the superconducting and normal-state properties under hydrostatic and
uniaxial pressure since they can be carried out on a single sample. \ Whereas
in transition-metal compounds the pressure-induced changes in $T_{c}$ can be
quite complicated, in simple metal superconductors like Al, In, Sn, or Pb,
$T_{c}$ invariably \textit{decreases} with increasing hydrostatic pressure
\cite{n13} due to lattice stiffening, electronic properties playing a
subordinate role.

In all high-pressure experiments on MgB$_{2}$ known to us
\cite{n14,monte,n14',n18,lorenz2,choi,tissen3}, $T_{c}$ is found to
monotonically decrease under pressure, but the magnitude of $dT_{c}/dP$ varies
considerably (see the Table below). \ Several studies have been carried out in
the pressure range to 2 GPa: \ in ac susceptibility measurements in a
piston-cylinder cell with liquid Fluorinert pressure medium, Lorenz \textit{et
al.} \cite{n14} obtain $dT_{c}/dP\simeq$ -1.6 K/GPa, whereas in electrical
resistivity studies with similar pressure techniques, Saito \textit{et al.}
\cite{n14'} and Choi \textit{et al.} \cite{choi} find -2.0 K/GPa and -1.36
K/GPa, respectively. \ Recent ac susceptibility \cite{n18} and
neutron-diffraction \cite{r10} measurements in a He-gas apparatus on a high
quality, isotopically pure ($^{11}$B) sample yield $dT_{c}/dP\simeq-1.11(2)$
K/GPa and the bulk modulus $B\simeq147.2(7)$ GPa, respectively, giving the
volume dependence $d\ln T_{c}/d\ln V=Bd\ln T_{c}/dP=+4.16(8);$ these authors
suggest that the lower magnitude of $dT_{c}/dP$ may result from the complete
lack of shear-stress effects in the helium pressure medium. \ Good agreement
with this result was obtained in subsequent He-gas studies by Lorenz
\textit{et al.} \cite{lorenz2} using an improved sample, where $dT_{c}%
/dP\simeq$ -1.07 K/GPa; a remeasurement in He-gas of the sample studied
earlier in Fluorinert yielded the comparable value $dT_{c}/dP\simeq$ -1.45
K/GPa, implying that differences among samples may affect $dT_{c}/dP$ as much
or more than shear stresses.

The first $T_{c}(P)$ studies in the pressure range above 2 GPa were carried
out by Monteverde \textit{et al.} \cite{monte} to 25 GPa in electrical
resistivity measurements using the solid pressure medium steatite in an
opposed anvil cell. \ These authors report widely differing pressure
dependences $T_{c}(P)$ for three of the four samples prepared, the initial
dependence $dT_{c}/dP$ varying from -0.35 to -0.8 K/GPa which they attribute
to lattice defects such as Mg nonstoichiometry. \ Very recently Tissen
\textit{et al.} \cite{tissen3} carried out ac susceptibility measurements in a
diamond-anvil-cell (DAC) to 28 GPa with methanol-ethanol pressure medium and
obtained the initial dependence $dT_{c}/dP\simeq-2$ K/GPa, finding that
$T_{c}$ decreases to 11 K at 20 GPa and 6 K at 28 GPa, almost twice the
decrease observed by Monteverde \textit{et al.} \cite{monte}. \ Tissen
\textit{et al.} \cite{tissen3} also report a broad bump in the pressure
dependence $T_{c}(P)$ near 9 GPa which they speculate arises from an
electronic Lifshitz transition. \ The results of all known $T_{c}(P)$
measurements on MgB$_{2}$ are summarized in the Table.

An excellent summary of high-pressure $T_{c}(P)$ data on MgB$_{2}$ and their
interpretation is contained in a recent paper by Chen \textit{et al.}
\cite{chen2}\textit{. \ }In the present paper we determine $T_{c}(P)$ in a DAC
with dense helium pressure medium and find $T_{c}$ to decrease monotonically
with pressure from 39.1 K to 20.9 K at 19.2 GPa. \ The initial pressure
dependence is in excellent agreement with that, $dT_{c}/dP\simeq-1.11(2)$
K/GPa, determined in earlier He-gas measurements to 0.7 GPa on the same
MgB$_{2}$ sample.

\section{EXPERIMENT}

The powder sample of MgB$_{2}$ for these studies is the same as used in
previous $T_{c}(P)$ \cite{n18} and neutron-diffraction \cite{r10} studies to
0.7 He-gas pressure and is made using isotopically-enriched $^{11}$B (Eagle
Picher, 98.46 atomic \% enrichment). \ A mixture of $^{11}$B powder (less than
200 mesh particle size) and chunks of Mg metal were reacted for 1.5 hours in a
capped BN crucible at 800$^{\circ}$C under an argon atmosphere of 50 bar.
\ The resulting sample displays sharp superconducting transitions in the ac
susceptibility with full shielding \cite{n18}; at ambient pressure the
temperature of the superconducting onset lies at 39.25 K and the midpoint at
39.10 K. \ In this paper we define $T_{c}$ at a given pressure from the
superconducting midpoint.

$T_{c}(P)$ can be determined to very high pressures using a helium-loaded
diamond-anvil-cell made of hardened Cu-Be alloy with binary Cu-Be inserts
fitted with 1/6-carat 16-sided type Ia diamond anvils (Drukker) and 0.5 mm
culet diameter. \ Gaskets of Ta$_{80}$W$_{20}$ alloy with diameter 2.7 mm and
thickness 290 $\mu m$ are preindented to about 70 $\mu m.$ \ The MgB$_{2} $
sample ($80\times80\times25$ $\mu$m$^{3}$) together with several small ruby
spheres (5-10 $\mu$m dia.) \cite{klotz} are placed in a 240 $\mu m$ dia. hole
drilled through the center of the gasket. \ The pressure clamp is placed in a
continuous flow cryostat and superfluid $^{4}$He is loaded into the gasket
hole at 2 K to serve as pressure medium. \ The pressure in the gasket hole can
be increased at any temperature at or below room temperature (RT) by loading a
double-membrane \cite{daniels} with a few bars of He gas to force the
diamond-anvils together. \ Temperature is measured by calibrated Pt and Ge
thermometers thermally anchored to the top diamond.

The pressure in the cell is determined to within 0.2 GPa at any temperature
below room temperature (RT) by measuring the pressure-induced shift in the
ruby R1 fluorescence line. \ A second ruby chip at ambient pressure located
just outside the pressure cell allows the correction for the temperature shift
of the ruby line. \ In the temperature range of this experiment (2 - 300 K)
the pressure-dependent shift in the wavelength of the ruby R1 fluorescence
line, $d\lambda/dP=3.642$ \AA/GPa, is independent of temperature \cite{ruby}.
\ Upon cooling from room temperature, there is negligible change in pressure
below 50 K; the pressure is normally measured at temperatures close to the
temperature of MgB$_{2}$'s superconducting transition.

In the DAC the superconducting transition of the MgB$_{2}$ sample is
determined inductively to $\pm$ 0.1 K using two balanced primary/secondary
coil systems connected to a Stanford Research SR830 digital lock-in amplifier
by slowly varying the temperature ($\sim$ 0.5 K/min.) through the transition.
\ The ac susceptibility studies were carried out using a 3 G (r.m.s.) magnetic
field at 1000 Hz. \ Over the transition the signal changed by $\sim$ 5 nV with
a background noise level of $\sim$ 0.2 nV, as seen in Fig. 1. \ Further
details of the He-gas and DAC high-pressure techniques are given elsewhere
\cite{r15}.

\section{RESULTS OF EXPERIMENT AND DISCUSSION}

In Fig. 1 we display the temperature dependence of the ac susceptibility at
three different pressures; the sharp superconducting transition is seen to
shift to lower temperatures with pressure. $\ T_{c}$ versus pressure is
plotted in Fig. 2 to 20 GPa, thus extending the pressure range of our previous
He-gas studies on the same MgB$_{2}$ sample nearly thirtyfold. \ Within
experimental error, $T_{c}$ decreases linearly with pressure to $\sim$ 10 GPa,
consistent with the rate -1.11 K/GPa (dashed line) from the earlier He-gas
studies \cite{n18}; the $T_{c}(P)$ data exhibit the upward (positive)
curvature expected from increasing lattice stiffening at higher pressures.
\ The decrease of $T_{c}$ with pressure is much less than that obtained by
Tissen \textit{et al.} \cite{tissen3}, but somewhat greater than the largest
dependence obtained by Monteverde \textit{et al.} \cite{monte}; in addition,
we do not observe the bump in $T_{c}(P)$ near 9 GPa reported by Tissen
\textit{et al.} \cite{tissen3}.

As seen in Fig. 2, the width of the superconducting transition generally
increases from $\sim$ 0.3 K for P $\leq10$ GPa to 0.9 K at 17.8 GPa,
increasing somewhat further for the data with decreasing pressure. \ This
increase in width $\Delta T_{c}$ is usually, but not always, accompanied by a
slight broadening of the ruby R1 fluorescence line; both broadening effects
point to a pressure gradient of approximately $\pm$ 0.3 GPa at the highest
pressures.\emph{\ }\ These broadening effects in $T_{c}$ are symptomatic of a
small ($\pm$ 1.5\%) pressure gradient in the cell arising from shear stresses
in the solid helium pressure medium. \ Since at $T_{c}\simeq$ 40 K helium is
fluid only for pressures $P\leq$ 0.5 GPa, all $T_{c}(P)$ data in Fig. 2 were
taken with the sample surrounded by solid helium. \ After the sample is cooled
through the melting curve of helium \cite{r14}, the differential thermal
contraction of solid helium, sample, and pressure cell lead to small shear
stresses and pressure gradients at temperatures near $T_{c}.$ \ In the
diamond-anvil-cell it is not possible to cool slowly through the melting curve
of helium with a well-defined temperature gradient, like in the He-gas
experiments \cite{n18}. \ In addition, since helium freezes at RT for
$P\approx$ 12 GPa, to increase the pressure above 12 GPa the diamonds must
compress solid helium, a process which adds further shear stresses. \ The
increase of the transition width with decreasing pressure (open circles in
Fig. 2) is curious and may indicate that the sample and ruby spheres have come
into direct contact with the gasket or diamond anvils.

At a given pressure, solid helium is the softest of all solids. \ Other
pressure media, such as frozen Fluorinert, methanol-ethanol or solid steatite,
will support much larger shear stresses and pressure gradients, with a
potential broadening and shifting of the superconducting transition,
particularly in elastically anisotropic samples such as MgB$_{2}$ where the
compressibility along the $c$ axis is 64\% larger than along the $a$ axis
\cite{r10}. \ Shear stresses are known to influence the pressure dependence of
$T_{c}$ in elastically anisotropic materials such as the high-$T_{c}$ oxides
\cite{n16} or organic superconductors \cite{n17}.

Since helium freezes for temperatures near $T_{c}\approx$ 40 K at 0.5 GPa, the
strictly linear pressure dependence of $T_{c}$ observed in the He-gas
measurements to 0.7 GPa \cite{n18} or 0.8 GPa \cite{lorenz2} shows that the
very weak shear stresses in solid He have negligible influence on the value of
$T_{c}$ in this material. \ The approximant agreement reported by Lorenz
\textit{et al.} \cite{lorenz2} for the pressure dependences of the same sample
in He-gas (-1.45 K/GPa) or frozen Fluorinert FC77 (-1.6 K/GPa) \cite{n14}
speaks against large shear stress effects in the latter pressure medium,
although the evidence here is not as compelling due to the absence of an error
estimate for $dT_{c}/dP$ and lack of comparative data on a series of samples.

Substantially larger shear stresses are generated in the frozen pressure
medium methanol-ethanol at very high pressures. \ At 20 GPa the width of the
superconducting transition in the ac susceptibility data of Tissen \textit{et
al.} \cite{tissen3} has increased by $\sim$ 3 K which corresponds to a
pressure gradient at that pressure of $\Delta P\approx3.5$ GPa or 18\%, an
order of magnitude higher than in the present helium-loaded DAC measurements;
a similar analysis of Tissen's data near 9 GPa yields a pressure gradient of
$\Delta P\approx1$ GPa. \ It is conceivable that the shear stresses leading to
these pressure gradients may be responsible for the bump in $T_{c}(P)$ near 9
GPa and perhaps also for the unusually large decrease in $T_{c}$ to 28 GPa
found by these authors \cite{tissen3}. \ Further experiments are required to
establish whether or not the anomalous $T_{c}(P)$ dependence is reproducible.
\ The same comment applies to the widely differing $T_{c}(P)$ results obtained
by Monteverde \textit{et al. } \cite{monte} in their quasi-hydrostatic
pressure measurements using the solid pressure medium steatite. \ In such a
pressure cell the shear stresses are often very large and sufficient to crush
dense samples or compact loosely sintered samples.

From a study of existing data, Tissen \textit{et al.} \cite{tissen3} have
suggested that larger values of $\left|  dT_{c}/dP\right|  $ are obtained for
samples exhibiting lower ambient-pressure values of $T_{c}$. \ The data in the
Table below lend some support to this suggestion. \ However, $T_{c}$ values
determined in ac susceptibility and electrical resistivity measurements are
difficult to compare directly, the latter usually lying higher. \ Further
experimentation under carefully controlled conditions is clearly necessary to
investigate this possible correlation.

To compare the present experimental results with theory, it is advantageous to
use the Murnaghan equation-of-state to convert the $T_{c}$ versus pressure $P$
data in Fig. 2 to $T_{c}$ versus relative volume $V/V_{0}$ data
\begin{equation}
\frac{V}{V_{0}}=\left[  1+\frac{B^{\prime}P}{B}\right]  ^{-1/B^{\prime}%
},\vspace{0.2cm}%
\end{equation}
where we use the value $B=147.2$ GPa from Ref. \cite{r10} and the canonical
value $B^{\prime}\equiv dB/dP=4$ supported by a recent calculation \cite{loa}.
\ The resulting dependence of $T_{c}$ on relative sample volume is shown in
Fig. 3$.$ \ The maximum pressure applied in the present experiment (19.2 GPa)
results in a volume decrease of $\sim$ 10\%.

We now compare the $T_{c}$ versus $V/V_{0}$ dependence in Fig. 3 to that from
our earlier He-gas data \cite{n18} on the same sample. \ As discussed in the
Introduction, in the latter study we obtained the initial volume dependence
$d\ln T_{c}/d\ln V=Bd\ln T_{c}/dP\simeq$ +4.16. \ Since the volume changes
from the 0.7 GPa He-gas pressure are very small ($\sim%
%TCIMACRO{\UNICODE{0xbd}}%
%BeginExpansion
\frac12
%EndExpansion
$ \%), we can write $d\ln T_{c}/d\ln V\simeq\lbrack\Delta T_{c}/T_{c0}%
]/[\Delta V/V_{0}]$ from which follows the linear relation
\begin{equation}
T_{c}\simeq T_{c0}\left(  -3.16+4.16\frac{V}{V_{0}}\right)
\end{equation}
which is plotted as the dashed line in Fig. 3. \ Note that the experimental
$T_{c}$ data at higher pressures now lie below the dashed line, the more so
the higher the pressure. \ The application of the equation of state has
transformed the superlinear $T_{c}$ versus $P$ data to sublinear $T_{c}$
versus $V/V_{0}$ data. \ The dashed line intercepts the horizontal axis
($T_{c}=$ 0 K) at $V/V_{0}=0.76$ which corresponds to a critical pressure of
$P_{c}\approx$ 75 GPa, an upper bound for the actual $P_{c}.$

Another method to extrapolate the initial volume dependence $d\ln T_{c}/d\ln
V\simeq$ +4.16 to higher pressures would be to assume that this differential
relation holds at all pressures, allowing us to integrate this expression to
obtain
\begin{equation}
T_{c}=T_{c0}\left(  \frac{V}{V_{0}}\right)  ^{+4.16},
\end{equation}
\vspace{0.1cm}which is plotted as the upper solid line in Fig. 3. \ This
volume dependence must, of course, agree with the He-gas data at low
pressures, where $V/V_{0}\simeq1,$ but rises well above the present
experimental data at higher pressures, yielding a much less satisfactory
agreement with the experimental data than the above linear approximation.
\ Note that according to Eq. (3) $T_{c}$ only reaches 0 K at infinite pressure
where $V\rightarrow0.$

It is not surprising that $T_{c}$ is not a linear function of $V/V_{0}$ to
very high pressures. \ The reason is that $T_{c}$ depends exponentially on
many fundamental parameters and it is the relatively small changes in these
parameters which lead to the large change in $T_{c}$ under pressure seen in
Figs. 2 and 3. \ Consider the McMillan equation \cite{r17}
\begin{equation}
T_{c}\simeq\frac{\left\langle \omega\right\rangle }{1.20}\exp\left\{
\frac{-1.04(1+\lambda)}{\lambda-\mu^{\ast}(1+0.62\lambda)}\right\}  ,
\end{equation}
valid for strong coupling ($\lambda\lesssim1.5),$ which connects the value of
$T_{c}$ with the electron-phonon coupling parameter $\lambda,$ an average
phonon frequency $\left\langle \omega\right\rangle ,$ and the Coulomb
repulsion $\mu^{\ast}$. \ The coupling parameter is defined by $\lambda
=N(E_{f})\left\langle I^{2}\right\rangle /[M\left\langle \omega^{2}%
\right\rangle ],$ where $N(E_{f})$ is the electronic density of states at the
Fermi energy, $\left\langle I^{2}\right\rangle $ the average squared
electronic matrix element, $M$ the molecular mass, and $\left\langle
\omega^{2}\right\rangle $ the average squared phonon frequency.

We now proceed to estimate the dependence of $T_{c}$ on relative volume
$V/V_{0}$ by inserting into the McMillan equation the relatively small volume
(pressure) dependences of each parameter $\left\langle \omega\right\rangle , $
$\lambda,$ and $\mu^{\ast}$ obtained from our previous analysis of the
high-pressure data on MgB$_{2}$ to 0.7 GPa \cite{n18}. \ In our previous
paper, we showed that the volume dependences $\gamma\equiv-d\ln\left\langle
\omega\right\rangle /d\ln V=$ +2.36 and $d\ln\lambda/d\ln V=+3.72$ gave a good
account of the experimental data to 0.7 GPa; $\mu^{\ast}=0.1$ was assumed to
be independent of pressure, a quite good assumption \cite{chen2}. \ As
suggested in the analysis of Chen \textit{et al.} \cite{chen2}, we now
integrate the volume dependences of these two parameters to obtain
\begin{equation}
\left\langle \omega\right\rangle =\left\langle \omega\right\rangle
_{0}(V/V_{0})^{-2.36}\text{ \ \ and \ \ \ }\lambda=\lambda_{0}(V/V_{0}%
)^{3.72}.
\end{equation}
For the initial values of the parameters we use the logarithmically averaged
phonon energy from inelastic neutron studies \cite{n9} $\left\langle
\omega\right\rangle _{0}=670$ K and $\lambda_{0}=0.90$ from the McMillan
equation. Inserting these two volume dependences in the McMillan equation, we
obtain the dependence of $T_{c}$ on relative volume shown as the lower solid
line in Fig. 3. \ The agreement with the experimental data is remarkable, much
better than in the two previous approximations.

According to this estimate, approximately 50 GPa hydrostatic pressure would be
required to drive $T_{c}$ to below 4 K. \ As in our earlier paper on the
He-gas results to 0.7 GPa \cite{n18}, the above analysis of the present
$T_{c}(P)$ data to 20 GPa demonstrates that the monotonic decrease of $T_{c}$
with pressure arises predominantly from the decrease in $\lambda$ due to
lattice stiffening ($d\ln\left\langle \omega^{2}\right\rangle /d\ln
V\simeq-2\gamma\simeq-4.72$), and not from electronic effects ($d\ln
[N(E_{f})\left\langle I^{2}\right\rangle ]/d\ln V\simeq-1$). A similar
calculation was very recently carried out by Chen \textit{et al.} \cite{chen2}
over a much wider pressure range; this paper also contains a detailed
discussion of the pressures dependences of all parameters, including
$\mu^{\ast}.$

The good agreement between the experimental data to 20 GPa and the predictions
of the McMillan formula using the volume dependences determined from the
He-gas data to 0.7 GPa lends additional evidence that superconductivity in
MgB$_{2}$ originates from standard BCS phonon-mediated electron
pairing.\vspace{0.4cm}

\noindent\textbf{Acknowledgments}

The authors are grateful to X. J. Chen for proving a preprint of his recent
paper. \ The authors would like to thank S. Klotz for providing the ruby
spheres. \ Work at Washington University supported by NSF grant DMR-0101809
and that at the Argonne National Laboratory by the U.S. Department of Energy,
Office of Science, contract No. W-31-109-ENG-38.\newpage

\begin{center}
{\large REFERENCES}

\bigskip
\end{center}

\noindent\textbf{Table. \ }Summary of available high-pressure $T_{c}(P)$ data
on MgB$_{2}$. \ $T_{c}$ values are at ambient pressure from superconducting
midpoint in ac susceptibility $\chi_{ac}$ and electrical resistivity $\rho$
measurements. \ $dT_{c}/dP$ is initial pressure derivative. \ $P^{\max}$(GPa)
is the maximum pressure reached in experiment .\vspace{0.2cm}

\begin{center}%
\begin{tabular}
[c]{|l|l|l|l|l|l|}\hline
$T_{c}$(K) & $\frac{dT_{c}}{dP}$(K/GPa) & $P^{\max}$(GPa) &
\textbf{measurement} & $%
\begin{array}
[c]{c}%
\text{\textbf{pressure}}\\
\text{\textbf{medium}}%
\end{array}
$ & \textbf{reference}\\\hline\hline
39.1 & -1.1 & 19.2 & $\chi_{ac}$, $^{11}$B isotope & helium & present
data\\\hline
39.1 & -1.11(2) & 0.66 & $\chi_{ac}$, $^{11}$B isotope & helium &
\cite{n18}\\\hline
39.1 & -1.09(4) & 0.63 & $\chi_{ac}$, $^{11}$B isotope & helium &
\cite{hamlin}\\\hline
39.2 & -1.11(3) & 0.61 & $\chi_{ac}$, $^{11}$B isotope & helium &
\cite{hamlin}\\\hline
40.5 & -1.12(3) & 0.64 & $\chi_{ac}$, $^{10}$B isotope & helium &
\cite{hamlin}\\\hline
39.2 & -1.07 & 0.84 & $\chi_{ac}$ & helium & \cite{lorenz2}\\\hline
37.4 & -1.45 & 0.84 & $\chi_{ac}$ & helium & \cite{lorenz2}\\\hline
37.4 & -1.6 & 1.84 & $\chi_{ac}$ & Fluorinert FC77 & \cite{n14}\\\hline
37.3 & -2 & 27.8 & $\chi_{ac}$ & 4:1 methanol-ethanol & \cite{tissen3}\\\hline
38.2 & -1.36 & 1.46 & $\rho$ & 1:1 daphne-kerosene & \cite{choi}\\\hline
37.5 & -1.9 & 1.35 & $\rho$ & Fluorinert FC70 & \cite{n14'}\\\hline
$\sim$ 35 & -0.35 to -0.8 & 25 & $\rho$ & steatite, RT solid & \cite{monte}%
\\\hline
\end{tabular}
\newpage\textbf{FIGURE \ CAPTIONS}

\bigskip
\end{center}

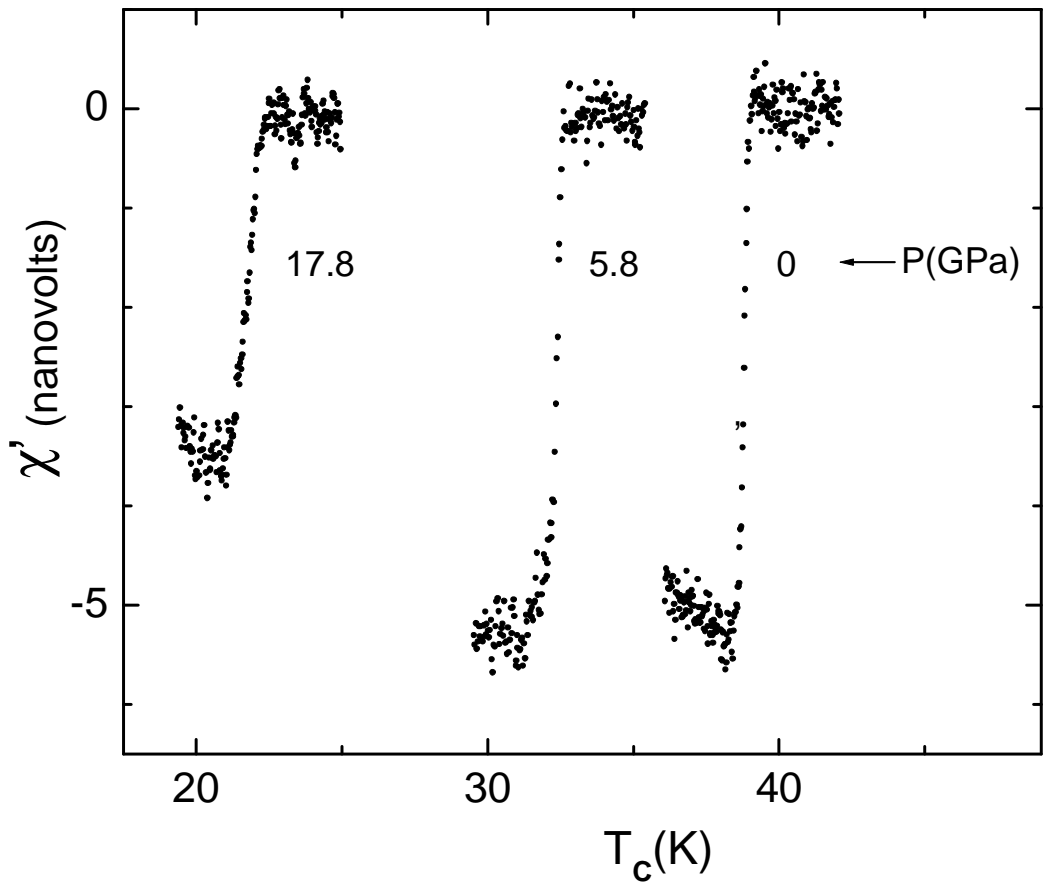
\noindent\textbf{Figure 1.} \ Real part of the ac susceptibility versus
temperature at three pressures.\bigskip

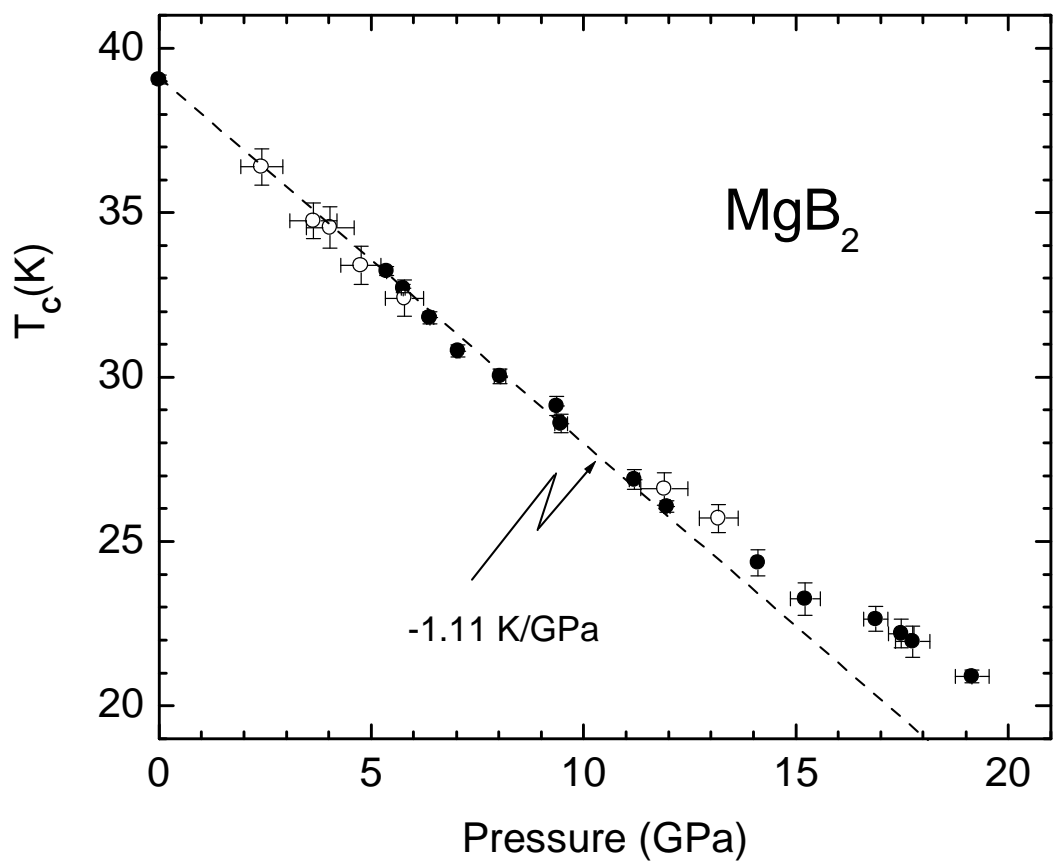
\noindent\textbf{Figure 2.} \ Superconducting transition temperature (at
midpoint) of MgB$_{2}$ versus pressure to 20 GPa from DAC measurements.
\ Vertical error bars give width of transition; horizontal error bars give
pressure gradient from broadening of ruby line. \ Data with filled circles
($\bullet)$ taken for monotonically increasing pressure, with open circles
($\circ$) for monotonically decreasing pressure. \ The straight dashed line
has slope -1.11 K/GPa.\bigskip

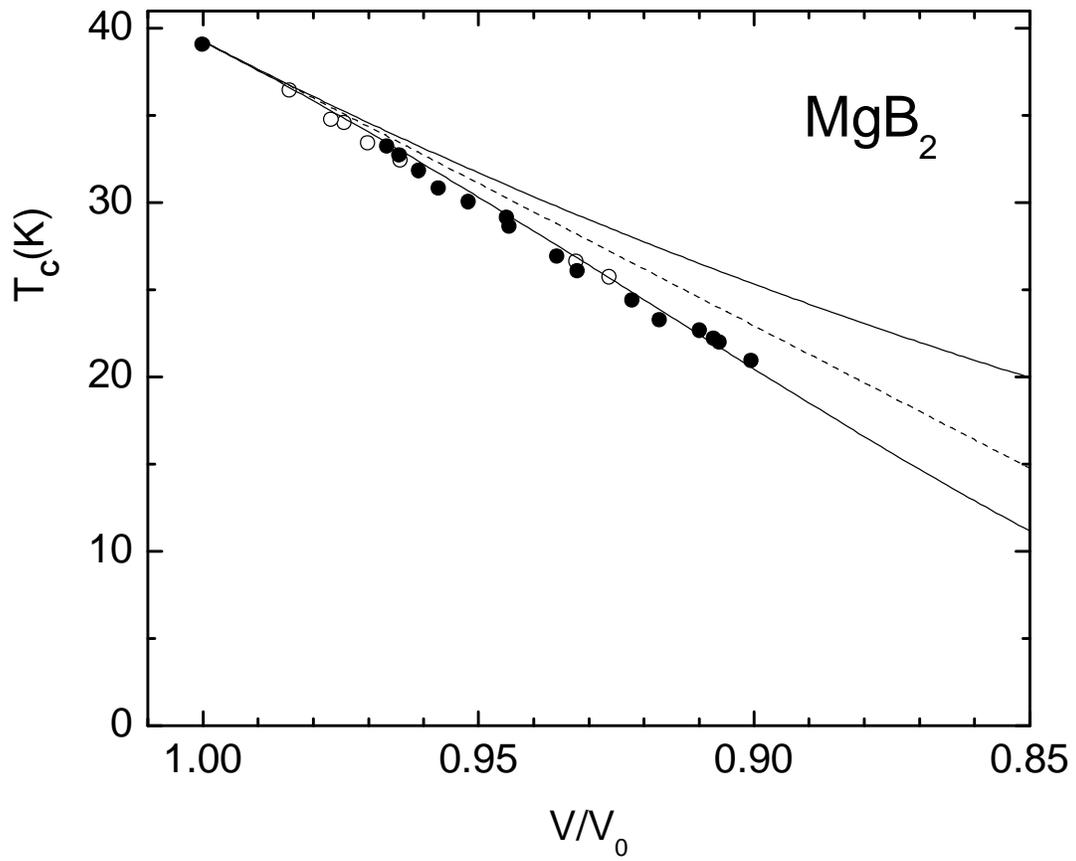
\noindent\textbf{Figure 3.} \ \ $T_{c}$ data from figure 2 plotted versus
relative volume $V/V_{0}$. \ See text for explanation of solid and dashed lines.

\begin{thebibliography}{99}
\bibitem{n4}J. Nagamatsu, N. Nakagawa, T. Muranaka, Y. Zenitani, and J.
Akimitsu, Nature 410 (2001) 63.

\bibitem {n13}See, for example, P. E. Seiden, Phys. Rev. 179 (1969) 458.

\bibitem {n14}B. Lorenz, R. L. Meng, C. W. Chu, Phys. Rev. B (in press);
preprint cond-mat/0102264.

\bibitem {monte}M. Monteverde, M. N\'{u}\~{n}ez-Regueiro, N. Rogado, K. A.
Regan, M. A. Hayward, T. He, S. M. Loureiro, R. J. Cava, Science 292 (2001) 75.

\bibitem {n14'}E. Saito, T. Taknenobu, T. Ito, Y. Iwasa, K. Prassides, and T.
Arima, J. Phys.: \ Condens. Matter 13 (2001) L267.

\bibitem {n18}T. Tomita, J. J. Hamlin, J. S. Schilling, D. G. Hinks, and J. D.
Jorgensen, Phys. Rev. B (in press); preprint cond-mat/0103538.

\bibitem {lorenz2}B. Lorenz, R. L. Meng, and C. W. Chu, preprint cond-mat/0104303.

\bibitem {choi}E. S. Choi, W. Kang, J. Y. Kim, Min-Seok Park, C. U. Jung,
Heon-Jung Kim, and Sung-Ik Lee, preprint cond-mat/0104454.

\bibitem {tissen3}V. G. Tissen, M. V. Nefedova, N. N. Kolesnikov, and M. P.
Kulakov, preprint cond-mat/0105475.

\bibitem {r10}J. D. Jorgensen, D. G. Hinks, S. Short, Phys. Rev. B 63 (2001) 224522.

\bibitem {chen2}X. J. Chen, H. Zhang, and H.-U. Habermeier, preprint.

\bibitem {klotz}The ruby spheres, which were annealed at 1500$^{\circ}$C for
one week followed by a cooldown to RT over a three-week period, were prepared
by J. C. Chervin, Physique des Milieux Condens\'{e}s, Paris, and contained
3600 ppm Cr$^{3+}.$ \ 

\bibitem {daniels}W. B. Daniels and W. Ryschkewitsch, Rev. Sci. Instrum. 54
(1983) 115.

\bibitem {ruby}R. A. Noack and W. B. Holzapfel, in: \ \textit{High Pressure
Science \& Technology, }edited by K. B. Timmerhaus and M. S. Barber (Plenum,
N.Y., 1979) Vol. 1, p. 748.

\bibitem {r15}J. S. Schilling, J. Diederichs, S. Klotz, R. Sieburger, in:
\ \textit{Magnetic Susceptibility of Superconductors and Other Spin Systems},
edited by R. A. Hein, T. L. Francavilla, D. H. Liebenberg (Plenum Press, New
York, 1991), p. 107.

\bibitem {r14}I. L. Spain and S. Segall, Cryogenics 11 (1971) 26.

\bibitem {n16}S. Klotz and J. S. Schilling, Physica C 209 (1993) 499; J. S.
Schilling and S. Klotz, in: \textit{Physical Properties of High Temperature
Superconductors}, Vol. III, ed. D. M. Ginsberg (World Scientific, Singapore,
1992) p. 59.

\bibitem {n17}T. Ishiguro, H. Ito, Y. Yamauchi, E. Ohmichi, M. Kubota, H.
Yamochi, G. Saito, M. V. Kartsovnik, M. A. Tanatar, Yu. V. Sushko, G. Yu.
Logvenov, Synthetic Metals 85 (1997) 1471; D. Jerome and H. J. Schulz, Adv.
Phys. 31 (1982) 299.

\bibitem {loa}I. Loa and K. Syassen, Solid State Commun. 118 (2001) 279.

\bibitem {r17}W. L. McMillan, Phys. Rev. 167 (1968) 331; P. B. Allen and R. C.
Dynes, Phys. Rev. B 1\textbf{2} (1975) 905.

\bibitem {n9}R. Osborn, E. A. Goremychkin, A. I. Kolesnikov, D. G. Hinks,
preprint cond-mat/0103064.

\bibitem {hamlin}J. J. Hamlin, C. Looney, T. Tomita, J. S. Schilling, D. G.
Hinks, and J. D. Jorgensen (unpublished).\newpage
\end{thebibliography}
\end{document}